\documentclass[journal=jpcafh,manuscript=article]{achemso}
\usepackage[T1]{fontenc}
\usepackage{graphicx}
\usepackage{amsmath,amssymb}
\usepackage[version=4]{mhchem}
\usepackage{xcolor}
\SectionNumbersOn

\newif\ifprov \provfalse

\newcommand{\BarSingO}{28.0}
\newcommand{\BarCrossO}{20.0}

\newcommand{\GapTopO}{\ensuremath{-}0.63}

\newcommand{\PhopO}{0.06}
\newcommand{\PsingleO}{0.03}

\newcommand{\OpenSeamO}{1.9}

\newcommand{\CuspO}{1.4}

\newcommand{\FballO}{0.12}

\newcommand{\FtunO}{0.97}
\newcommand{\ShareBelowO}{99.5}

\newcommand{\KappaTunO}{59}
\newcommand{\HalfTunO}{240~ms}

\newcommand{\KappaWkbO}{260}

\newcommand{\HalfBestO}{14~s}

\newcommand{\RotGainO}{39}
\newcommand{\SpeedupO}{\ensuremath{2.0\times10^{3}}}
\newcommand{\AtwoTopO}{\ensuremath{-}0.36}

\newcommand{\AtwoSeamO}{22.6}
\newcommand{\VrotO}{16}
\newcommand{\ProtO}{\ensuremath{2.0\times10^{-3}}}
\newcommand{\PhopOverProtO}{29}

\newcommand{\KappaRotO}{41}

\newcommand{\TunLeadO}{\ensuremath{2.6\times10^{3}}}

\newcommand{\RefShiftLoO}{1.6}
\newcommand{\RefShiftHiO}{2.2}
\newcommand{\VsoVrotO}{6}

\newcommand{\VcmCtlO}{87}
\newcommand{\BarSingS}{35.4}

\newcommand{\PhopS}{0.49}
\newcommand{\PsingleS}{0.31}
\newcommand{\BypassS}{9.0}

\newcommand{\OpenSeamS}{20.2}

\newcommand{\FballS}{0.30}

\newcommand{\BarOpenS}{36.2}

\newcommand{\RotGainS}{41}
\newcommand{\SpeedupS}{\ensuremath{5.3\times10^{4}}}

\newcommand{\AtwoSeamS}{29.2}

\newcommand{\RefRingOpenNevS}{\ensuremath{+}0.8}
\newcommand{\RefRingOpenCCS}{\ensuremath{+}5.2}

\newcommand{\RefShiftLoS}{4.0}
\newcommand{\RefShiftHiS}{4.5}

\newcommand{\VcmCtlS}{199}

\newcommand{\ExtraSe}{2.6}

\newcommand{\ThetaCrossSe}{76.9}

\newcommand{\BarOpenSe}{24.3}

\newcommand{\RotGainSe}{6.8}
\newcommand{\SpeedupSe}{\ensuremath{6.2\times10^{6}}}

\newcommand{\AtwoSeamSe}{23.8}

\newcommand{\RefShiftLoSe}{1.0}
\newcommand{\RefShiftHiSe}{1.9}

\newcommand{\CtlLSSe}{20.7}
\newcommand{\CtlFullSe}{18.5}
\newcommand{\CtlParSe}{18.5}
\newcommand{\CtlPerpSe}{20.5}
\newcommand{\CtlGainFullSe}{2.2}
\newcommand{\CtlGainParSe}{2.2}
\newcommand{\CtlGainPerpSe}{0.2}

\newcommand{\VzetaSe}{0.57}

\newcommand{\ExtraTe}{5.5}

\newcommand{\BarOpenTe}{16.6}

\newcommand{\SpeedupTe}{\ensuremath{3.5\times10^{8}}}

\newcommand{\AtwoSeamTe}{21.5}

\newcommand{\CtlLSTe}{18.8}
\newcommand{\CtlFullTe}{13.6}
\newcommand{\CtlParTe}{13.7}
\newcommand{\CtlPerpTe}{17.5}
\newcommand{\CtlGainFullTe}{5.2}
\newcommand{\CtlGainParTe}{5.1}
\newcommand{\CtlGainPerpTe}{1.3}

\newcommand{\BarSingPo}{23.2}

\newcommand{\BarSOPo}{3.6}

\newcommand{\RingOpenPo}{\ensuremath{-}9.2}
\newcommand{\RingOpenSOPo}{\ensuremath{-}1.6}
\newcommand{\GapTopPo}{\ensuremath{-}0.29}

\newcommand{\BypassPo}{6.5}
\newcommand{\ExtraPo}{13.1}

\newcommand{\SensRefPo}{3.5}
\newcommand{\SensThreePo}{3.4}
\newcommand{\SensNinePo}{3.5}

\newcommand{\SensExpPo}{2.3}

\newcommand{\HalfBestPo}{59~ps}

\newcommand{\HalfSfPo}{32~min}

\newcommand{\SpeedupPo}{\ensuremath{3.3\times10^{13}}}
\newcommand{\AtwoTopPo}{\ensuremath{-}0.18}

\newcommand{\AtwoSeamPo}{19.0}
\newcommand{\VrotPo}{2.1}

\newcommand{\VsoVrotPo}{\ensuremath{2.8\times10^{3}}}
\newcommand{\CtlLSPo}{16.2}
\newcommand{\CtlFullPo}{3.5}
\newcommand{\CtlParPo}{2.5}
\newcommand{\CtlPerpPo}{8.5}

\newcommand{\ScanGapO}{173}

\newcommand{\ScanHalfO}{0.25}
\newcommand{\ScanGapS}{397}

\newcommand{\ScanHalfS}{0.57}

\newcommand{\SadVminSe}{0.07}

\newcommand{\SadGapMin}{0.02}
\newcommand{\SadGapMax}{0.08}

\newcommand{\OrbGapO}{0.013}

\newcommand{\OrbGapS}{0.031}
\newcommand{\OrbCrossSe}{89}

\newcommand{\OrbGapSe}{0.14}

\newcommand{\OrbGapTe}{0.28}

\newcommand{\OrbGapPo}{0.67}
\newcommand{\OrbLamZetaMin}{0.69}
\newcommand{\OrbLamZetaMax}{0.71}

\newcommand{\EfourRatioMin}{0.07}
\newcommand{\EfourRatioMax}{0.18}
\newcommand{\ProtGapP}{0.029}

\newcommand{\ProtGapBi}{0.92}

\newcommand{\ProtGapZetaMin}{0.89}
\newcommand{\ProtGapZetaMax}{1.01}
\newcommand{\OneAtomDevMax}{0.4}

\newcommand{\OneNobleDevMax}{0.4}
\newcommand{\OneNobleLast}{Rn}

\newcommand{\CasTopAbsSe}{0.5}

\newcommand{\CasTopAbsTe}{1.1}

\newcommand{\CasTopAbsPo}{1.1}
\newcommand{\CasBarEstPo}{2.4}

\newcommand{\NDiscPo}{2}
\newcommand{\NDiscarded}{2}

\newcommand{\VzetaMin}{0.55}
\newcommand{\VzetaMax}{0.58}

\newcommand{\CtlParDevMax}{0.2}

\newcommand{\CtlRingOpenS}{\ensuremath{+}0.7}

\newcommand{\CtlParDevSe}{0.02}
\newcommand{\CtlParDevTe}{0.19}

\newcommand{\ChkQzBarMax}{1.3}
\newcommand{\ChkQzDiffMax}{0.7}

\newcommand{\ChkQzParMax}{0.2}

\newcommand{\ChkQzFullPo}{4.1}

\newcommand{\ChkFvDiffMax}{0.4}

\newcommand{\ChkFvRingOpenS}{\ensuremath{+}3.8}

\newcommand{\ChkFvBarShiftAbsS}{2.8}

\newcommand{\ChkFvFullPo}{3.2}

\newcommand{\kcal}{kcal\,mol$^{-1}$}
\newcommand{\cm}{cm$^{-1}$}
\newcommand{\Hso}{\hat H_{\mathrm{SO}}}


\author{Shadan Ghassemi Tabrizi}
\email{s.ghassemi-tabrizi@hzdr.de}
\affiliation{Computational System Sciences, Technische Universit\"at Dresden, 01187 Dresden, Germany}
\alsoaffiliation{Center for Advanced Systems Understanding (CASUS), Am Untermarkt 20, 02826
G\"orlitz, Germany}

\title{Spin--Orbit Coupling in Symmetry-Forbidden Reactions: A Selection Rule Tested on the
Chalcogen Trimers}

\AtBeginDocument{\singlespacing}
\begin{document}

\begin{abstract}
A reaction is forbidden by orbital symmetry when an occupied and an empty orbital of different
symmetry change places along its path. With spin--orbit coupling orbitals become spinors, and
those of two such orbitals can mix only if the orbitals differ by the symmetry of a rotation;
the operator component of that symmetry couples ground state and diradical triplet. We test this rule on the ring
opening of the chalcogen trimers \ce{O3} to \ce{Po3}, where an in-plane and an out-of-plane orbital
cross ($C_{2v}$ surfaces; NEVPT2, spin--orbit configuration interaction),
by letting the operator components act separately. In \ce{Se3} and \ce{Te3} the
component named by the rule alone lowers the barrier by \CtlGainParSe{} and \CtlGainParTe{} \kcal{}
(complete operator: \CtlGainFullSe{} and \CtlGainFullTe),
the other two by \CtlGainPerpSe{} and \CtlGainPerpTe{} \kcal; in \ce{Po3} they
contribute in second order. The triplet crosses the ground state \BypassPo{} to \BypassS{} \kcal{}
below the forbidden saddle. Spin--orbit coupling makes this crossing passable in \ce{O3} and
\ce{S3} and, from \ce{Se3} on, merges the two states into one surface whose saddle lies \ExtraSe,
\ExtraTe, and \ExtraPo{} \kcal{} lower. Molecular rotation obeys the same rule but is
\VsoVrotO{} to \VsoVrotPo{} times weaker, which bears on the lifetime predicted for cyclic
ozone.
\end{abstract}

\section{Introduction}

The Woodward--Hoffmann rules\cite{WH1965,LH1965,WH1969} relate the course of a concerted reaction to
the symmetry of a few orbitals. A reaction is called forbidden when, on the way from reactant to
product, an occupied and an empty orbital of different symmetry have to change places. Because the
two cannot mix, the energy rises along the path until the molecule has the character of a diradical.
How high this point lies and how molecules avoid it has been studied for sixty years, and the
question has recently been taken up again.\cite{Zhou2024}

The rules were formulated without spin--orbit coupling: their orbitals are functions of space alone
and carry the labels of the ordinary point group. With spin--orbit coupling the one-electron
functions are spinors, which are classified by the double-valued irreducible representations of the
double group,\cite{Bethe1929,Altmann1994} and there are fewer of these. In $C_{2v}$ there is only
one, so that all spinors of a $C_{2v}$ molecule have the same symmetry and, formally, no crossing of
two of them is protected. Where there is a center of inversion, on the other hand, \emph{gerade} and
\emph{ungerade} spinors remain distinct. The effect of spin--orbit coupling on an orbital-symmetry
rule therefore depends on the symmetries involved: some forbidden crossings can become avoided,
whereas others remain strictly forbidden. In this paper we ask which crossings belong to which
group and how large the effect becomes as the atoms get heavier.

The first question is answered by a selection rule. The spinors that derive from two orbitals $a$
and $b$ have a symmetry in common if the product $\Gamma_a\otimes\Gamma_b$ is totally symmetric or
the representation of a rotation, and only then. For a forbidden crossing the first case does not
arise, which leaves the second: \emph{spin--orbit coupling can act on a forbidden crossing if, and
only if, the two crossing orbitals differ by the symmetry of a rotation}. This can be understood
from the form of the operator. Its spatial part is an angular momentum and therefore connects two
orbitals that a rotation converts into each other. The state to which it couples the ground state
is the triplet of the diradical configuration. At the crossing the two electrons occupy two
orbitals in the same region of space, and by Hund's rule this triplet lies below the corresponding
singlet\cite{Salem1972,Borden1977} and, in a two-orbital model, below the forbidden barrier
(Section~\ref{sec:states}). The symmetry by which $a$ and $b$ differ, which makes the reaction
forbidden, thus also determines which component of the spin--orbit operator, if any, gives access
to this lower-lying state. Such a rule applies only as long as the path retains the symmetry
that distinguishes the two orbitals, and it states whether a coupling is allowed, not how large it
is.

The elements of this argument are known. That a state of another spin can carry a reaction around a
barrier is the idea of two-state reactivity\cite{Schroder2000} and of the thermal chemiexcitation of
dioxetanes,\cite{Turro1973} and the rearrangement of Dewar benzenes to triplet benzene was described
as a spin-forbidden, nonadiabatic pericyclic reaction in 1973.\cite{Lechtken1973} The kinetics of
such reactions is described by nonadiabatic transition-state
theory,\cite{Lorquet1988,Harvey2007,Yang2018} and the rules for the size of spin--orbit coupling in
diradicals go back to Salem and Rowland\cite{Salem1972} and to Michl and
Havlas.\cite{Michl1996,Michl1997} We are not aware of a statement that the symmetry by which the
two crossing orbitals of a forbidden reaction differ determines whether, and through which of its
components, spin--orbit coupling can reach the diradical triplet, nor of a study that follows this
connection quantitatively through one reaction.

To test the rule we looked for a reaction in which the effect of spin--orbit coupling is not masked
by other ways around the barrier. In most forbidden reactions a vibration of symmetry
$\Gamma_a\otimes\Gamma_b$ mixes the two orbitals, and the molecule leaves the symmetric path. A
triatomic molecule does not have this possibility: it is planar in every geometry, and no vibration
mixes its in-plane and out-of-plane orbitals. The ring opening of ozone is a
well-studied forbidden reaction of this kind;\cite{Hay1977,Xantheas1991,Ivanic1997,Theis2016} as
Flemmig, Wolczanski, and Hoffmann pointed out, it is forbidden because the mirror plane that
separates $\sigma$ from $\pi$ orbitals is conserved along the whole path.\cite{Flemmig2005} Its
heavier congeners \ce{S3}, \ce{Se3}, \ce{Te3}, and \ce{Po3} undergo the same reaction with an atomic
spin--orbit constant that grows by a factor of about eighty,\cite{NIST2024} so that the strength of
spin--orbit coupling can be varied while the reaction stays the same. Three atoms are also few
enough for potential energy surfaces in both coordinates of the symmetric path, a multireference
treatment, a spin--orbit configuration interaction without truncation, and a calculation in which
the components of the spin--orbit operator are switched on one at a time.

Cyclic ozone has long been sought. Chen and Hu
found that the open-shell singlet of the diradical configuration crosses the ground state on the
ring-opening path and predicted, with multidimensional tunneling, that the ring survives for only a
fraction of a second at room temperature;\cite{Chen2011} at the transition-state geometry of a
CASSCF calculation, which lies near a crossing of singlet and triplet, a neural-network wave
function without symmetry constraints converged to a nearly pure triplet
($\langle\hat S^2\rangle=1.85$).\cite{Jiang2025} In the analysis presented here these two open-shell
states are the two partners that the selection rule admits, one for the rotation of the molecule
and one for spin--orbit coupling. The heavier trimers have been studied as
well,\cite{Basch1989,Balasubramanian1993,Goddard1999,Peterson2006} including the path between ring
and open form of \ce{S3}, \ce{Se3}, and \ce{Te3},\cite{Goddard1999} with the result that ring and
open form approach each other in energy and that the barrier between them falls down the group; for
\ce{Po3} a density-functional study with relativistic effective core potentials reported two bent
structures of different multiplicity and nearly the same energy.\cite{VanYperen2015} The barrier
between ring and open form in the presence of spin--orbit coupling has, to our knowledge, not been
examined.

\section{Symmetry Analysis}
\label{sec:theory}

\subsection{Orbital Crossing in the Ring Opening}
\label{sec:crossing}

We place the trimer \ce{E3} in the $yz$ plane with its twofold axis along $z$ and describe it by the
length $r$ of its two equal bonds and by the apex angle $\theta$, which is $60^\circ$ in the ring
and about $115^\circ$ in the open form. The symmetry is $C_{2v}$ throughout, and we use the orbital
labels of ozone for all five molecules.

\begin{figure}
\centering
\includegraphics{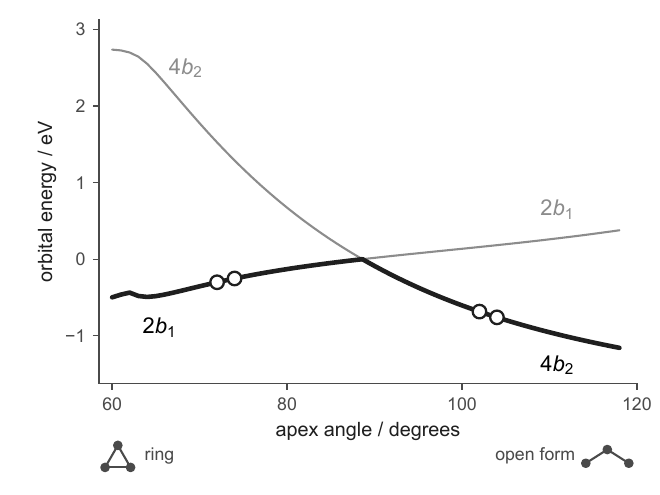}
\caption{The orbital crossing that forbids the ring opening, shown for \ce{Se3}: energies of the
out-of-plane orbital $2b_1$ and of the in-plane orbital $4b_2$ along a path from the ring to the open
form, relative to the point where they cross. Heavy line: the orbital that holds the two electrons
(circles) of the closed-shell ground state; thin line: the empty one.}
\label{fig:crossing}
\end{figure}

Figure~\ref{fig:crossing} shows the two orbitals involved when the ring of
\ce{Se3} is opened. The orbital $2b_1$ is the antibonding combination of the out-of-plane $p$
functions, a $\pi^*$ orbital. It is occupied in the ring, and its energy rises slowly as the angle
opens. The orbital $4b_2$ lies in the plane. In the ring it is the antibonding $\sigma^*$ orbital of
the bond that is broken, and its energy falls steeply as the two atoms move apart. Near
$\OrbCrossSe^\circ$ the two change places. One of them is antisymmetric and the other symmetric with
respect to the molecular plane, so that they cannot mix, and they cross. The ring has the
configuration $\dots(2b_1)^2(4b_2)^0$ and the open form $\dots(4b_2)^2(2b_1)^0$: two electrons have
to move from one orbital to the other, which is what characterizes a forbidden reaction.

The orbital energies of Figure~\ref{fig:crossing} belong to an operator that favors neither side,
the Fock operator of a density in which each of the two orbitals holds one electron
(Section~\ref{sec:methods}). In the many-electron states the changeover comes a few degrees
earlier, near $82^\circ$, because the two configurations differ in their electron repulsion. The
same two orbitals cross in all five trimers.

\subsection{Electronic States at the Crossing}
\label{sec:states}

Where two orbitals $a$ and $b$ cross, two electrons can be distributed over them in four ways. Two
are closed shells, $\Phi_a=|\dots a\bar a|$ for the ring and $\Phi_b=|\dots b\bar b|$ for the open
form. The other two have one electron in each orbital, coupled to a triplet $^3\Psi_{ab}$ or to a
singlet $^1\Psi_{ab}$; both have the spatial symmetry $\Gamma_a\otimes\Gamma_b$, here
$b_1\otimes b_2=a_2$, and are the states $^3A_2$ and $^1A_2$ of the trimers.

\begin{figure}
\centering
\includegraphics[width=\textwidth]{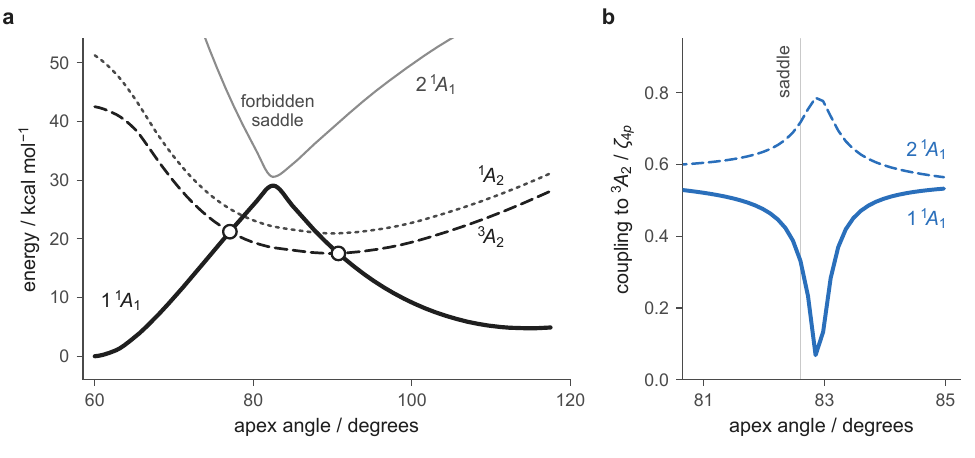}
\caption{The states at the crossing, shown for \ce{Se3} (NEVPT2). (a) Along the bending path, with
the bond length that minimizes $1\,^1A_1$; the lines join points computed every $2.5^\circ$. The
two $^1A_1$ states are the combinations of the two closed shells; their avoided crossing is the
forbidden saddle. The triplet $^3A_2$ (dashed) and the open-shell singlet $^1A_2$ (dotted) of the
diradical configuration $(4b_2)^1(2b_1)^1$ pass below it. Circles: the two crossings of the ground
state with the triplet. (b) Spin--orbit coupling of the two $^1A_1$ states to $^3A_2$ through the
saddle, in units of the atomic spin--orbit constant (scan in steps of $0.125^\circ$ at the bond
length of the saddle). Grey line: highest point of $1\,^1A_1$ on the scan.}
\label{fig:states}
\end{figure}

The two closed shells have the same symmetry and interact through the exchange integral
$K_{ab}=(ab|ab)$. Where their energies are equal they combine to $(\Phi_a\pm\Phi_b)/\sqrt2$, and the
lower combination, $\Psi_-$, is the ground state at the top of the forbidden barrier. The barrier is
therefore finite and sharply peaked: the two $^1A_1$ states of
Figure~\ref{fig:states} approach each other and repel.

The positions of the open-shell states follow from the same model. With the energies
$E_a=2h_a+J_{aa}$ and $E_b=2h_b+J_{bb}$ of the closed shells and $h_a+h_b+J_{ab}\mp K_{ab}$ of the
triplet and the open-shell singlet, one obtains at the point where $E_a=E_b$
\begin{equation}
E(^3\Psi_{ab})-E(\Psi_-)=-2K_{+-},\qquad E(^1\Psi_{ab})-E(\Psi_-)=2\,(K_{ab}-K_{+-}) ,
\label{eq:states}
\end{equation}
where $K_{+-}=\tfrac14(\rho_a-\rho_b\,|\,\rho_a-\rho_b)$ is the exchange integral of the orbitals
$(a\pm b)/\sqrt2$, the self-repulsion of the difference of the two orbital densities, which cannot
be negative. In this model the triplet therefore does not lie above the forbidden saddle; this is
Hund's rule for two orbitals that occupy the same region of space.\cite{Borden1977} The open-shell
singlet lies $2K_{ab}$ above the triplet. For two perpendicular $p$ functions on one atom
$K_{ab}=K_{+-}$, and $\Psi_-$ and $^1\Psi_{ab}$ are then two components of an atomic $^1D$ term. The
open-shell singlet is therefore expected close to the saddle, and below it when the two orbitals
are distributed differently over the atoms.

The calculations agree with this picture (Figure~\ref{fig:states}a). At the grid point nearest to
the saddle of the closed-shell singlet the triplet lies between \GapTopO{} (\ce{O3}) and
\GapTopPo{} eV (\ce{Po3}) relative to it and the open-shell singlet between \AtwoTopO{} and
\AtwoTopPo{} eV (Table~S21). Both diradical states thus pass below the forbidden
saddle in all five molecules. On the ring side of the saddle the ground state crosses first the
triplet and then the open-shell singlet, and on the far side it crosses them again. Whether these
crossings provide a route for the reaction depends on whether the molecule can change from one
state to the other.

\subsection{Coupling Operators and the Selection Rule}
\label{sec:turn}

Both diradical states differ from the ground state by the symmetry $a_2$, and a change of state
requires a perturbation of this symmetry. Three kinds of perturbation come into question.

The first is a vibration: a distortion of symmetry $\Gamma_a\otimes\Gamma_b$ mixes the open-shell
singlet into the ground state, and most forbidden reactions leave the symmetric path in this way. The three vibrations of a triatomic molecule have
the symmetries $a_1$, $a_1$, and $b_2$, so that this possibility does not exist here.

The second is the rotation of the molecule. The rotational energy
$\sum_kB_k(\hat J_k-\hat L_k)^2$ contains the terms $-2B_k\hat J_k\hat L_k$, which couple the
rotation of the nuclear frame to the orbital angular momentum $\hat L_k=\sum_i\hat l_k(i)$ of the
electrons, and
\begin{equation}
\langle\Phi_a|\hat L_k|{}^1\Psi_{ab}\rangle=-\langle\Phi_b|\hat L_k|{}^1\Psi_{ab}\rangle
=\sqrt2\,\langle a|\hat l_k|b\rangle .
\label{eq:rot}
\end{equation}
The rotation about the twofold axis has the symmetry $a_2$, so that in a rotating trimer the ground
state is coupled to the open-shell singlet.

The third is spin--orbit coupling. We write the operator as
$\Hso=\sum_i\hat{\mathbf h}(i)\cdot\hat{\mathbf s}(i)$, where $\hat{\mathbf h}$ is purely imaginary
and transforms like an axial vector; in the familiar one-center form
$\hat{\mathbf h}=\sum_A\zeta_A(r_A)\,\hat{\mathbf l}_A$. With
$\langle a|\hat h_k|b\rangle=i\lambda_k$ the component $M_S=0$ of the triplet, quantized along $k$,
has the matrix elements
\begin{equation}
\langle\Phi_a|\Hso|{}^3\Psi_{ab}\rangle=\langle\Phi_b|\Hso|{}^3\Psi_{ab}\rangle=-\frac{i\lambda_k}{\sqrt2} .
\label{eq:me}
\end{equation}
The component of $\hat{\mathbf h}$ along the twofold axis has the symmetry $a_2$ and connects the
ground state with $^3A_2$. The other two components, of symmetry $b_1$ and $b_2$, couple it to
other triplets, which do not arise from the two crossing orbitals.

\begin{figure}
\centering
\includegraphics{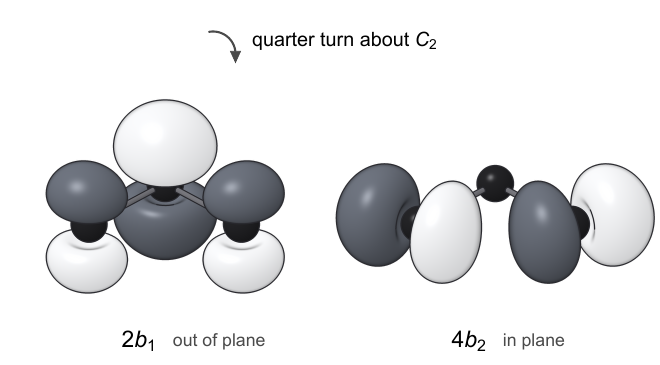}
\caption{The two crossing orbitals of \ce{Se3} near the crossing ($\theta=82.5^\circ$): surfaces of
constant value $\pm0.055$ $a_0^{-3/2}$, one phase shaded and the other light. The view is along the
twofold axis from the side of the two terminal atoms, raised by $28^\circ$ out of the molecular
plane, so that the apex atom lies behind and above them. On each terminal atom a quarter turn about
the axis converts the $p$ function of one orbital into that of the other; the apex atom contributes
to $2b_1$ only.}
\label{fig:orbitals}
\end{figure}

Both couplings contain the same quantity, the matrix element of an orbital angular momentum about
the twofold axis between the two crossing orbitals. Figure~\ref{fig:orbitals} illustrates why it is
large. Seen along the axis, $2b_1$ consists mainly of two $p$ functions that are perpendicular to
the molecular plane on the terminal atoms and $4b_2$ of two that lie in the plane. A quarter turn
about the axis converts one into the other, and this is the action of $\hat l_z$ on a $p$ function.
The situation corresponds to El-Sayed's rule for intersystem crossing.\cite{ElSayed1963} The
argument is a local one: it concerns the $p$ functions on the two terminal atoms, to which both
orbitals contribute; the apex atom contributes to $2b_1$ only.

In general terms, the matrix elements of eqs~\ref{eq:rot} and \ref{eq:me} vanish unless
\begin{equation}
\Gamma_a\otimes\Gamma_b\ni\Gamma(R_k) ,
\label{eq:rule}
\end{equation}
that is, unless the two orbitals differ by the symmetry of a rotation. This is the selection rule.
For spin--orbit coupling it also follows without perturbation theory. In the double group the two
spin functions span a double-valued representation $D^{1/2}$, and the spinors derived from an orbital
$a$ span $\Gamma_a\otimes D^{1/2}$. Those of $a$ and $b$ have a representation in common if and only
if $\Gamma_a\otimes\Gamma_b$ contains one of the representations in
\begin{equation}
D^{1/2}\otimes D^{1/2\,*}=\Gamma_1\oplus\Gamma(R_x)\oplus\Gamma(R_y)\oplus\Gamma(R_z) ,
\label{eq:dg}
\end{equation}
the totally symmetric one or one of the three rotations; the spin is an axial vector. If the product
of two crossing orbitals is none of these, as for a \emph{gerade} and an \emph{ungerade} orbital,
their spinors differ in symmetry and the crossing remains; the diradical triplet
then has no component of the symmetry of the ground state, and spin--orbit coupling cannot connect
the two in any order. The two closed shells still mix through $K_{ab}$, but there is no route
through the triplet. Diatomic molecules provide a familiar example: with spin--orbit
coupling the levels that derive from a $\sigma_g$ and a $\pi_g$ orbital avoid each other, whereas
those from a $\sigma_g$ and a $\pi_u$ orbital cross. If the product is a rotation, the crossing is
no longer protected by symmetry. How large the effect is then depends on the matrix element, about
which two points are needed below.

First, for a ground state $\Psi_0=c_a\Phi_a-c_b\Phi_b$ the two couplings are
\begin{equation}
V=|\langle\Psi_0|\Hso|{}^3\Psi_{ab}\rangle|=\frac{|\lambda_k|}{\sqrt2}\,|c_a-c_b| ,\qquad
V_{\mathrm{rot}}=2\sqrt2\,B_k|J_k|\,|\langle a|\hat l_k|b\rangle|\,(c_a+c_b) .
\label{eq:V}
\end{equation}
The spin--orbit coupling vanishes for the symmetric combination $\Psi_-$; this is the result of
Salem and Rowland that a purely covalent singlet diradical does not couple to its
triplet.\cite{Salem1972,Michl1996} In the model the coupling is thus allowed by symmetry at this
point and nevertheless zero. It is largest where one configuration dominates, and this is where
the crossings with the triplet lie, on either side of the saddle (Figure~\ref{fig:states}a). The
computed coupling behaves accordingly (Figure~\ref{fig:states}b). In \ce{Se3} the ground state
meets the triplet at $\ThetaCrossSe^\circ$, where the configuration of the ring still dominates and
the coupling is \VzetaSe{} $\zeta_{np}$. Next to the highest point of the ground state, at
$82.60^\circ$, the coupling passes through a minimum of \SadVminSe{} $\zeta_{np}$ at
$82.85^\circ$, close to the angle at which the two crossing orbitals are equally occupied, while
the upper $^1A_1$ state takes over the coupling. The other four molecules behave in the same way, with minima of 0.02 to
0.12 $\zeta_{np}$ within $0.35^\circ$ of the saddle (Figure~S3,
Table~S11); the exact zero is a property of the model. Second, the two couplings
contain
the same orbital matrix element, multiplied in one case by the spin--orbit constant of the atom and
in the other by the rotational energy of the molecule. For
$c_a\gg c_b$ and orbitals built from the $p$ functions of one kind of atom,
\begin{equation}
\frac{V}{V_{\mathrm{rot}}}\approx\frac{\zeta}{4B_k|J_k|} .
\label{eq:ratio}
\end{equation}
Spin--orbit coupling thus plays the same role for the triplet as rotation does for the open-shell
singlet.

\subsection{Correlation Diagram with Spin--Orbit Coupling}
\label{sec:spinors}

\begin{figure}
\centering
\includegraphics[width=\textwidth]{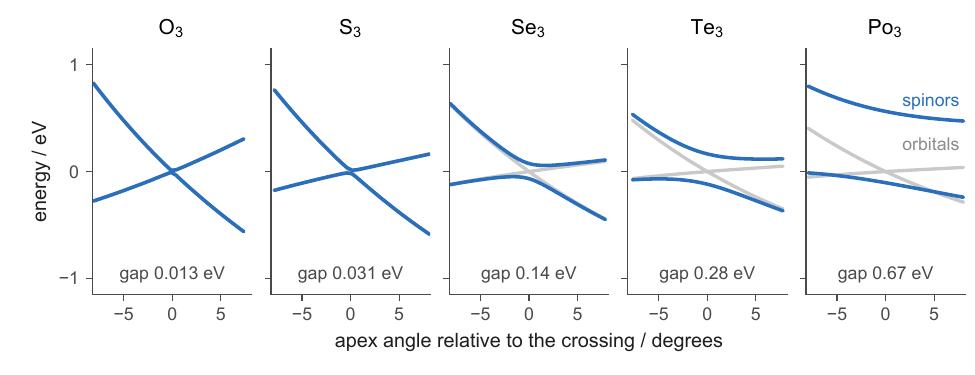}
\caption{The correlation diagram with spin--orbit coupling. Grey: the energies of the orbitals
$2b_1$ and $4b_2$ of Figure~\ref{fig:crossing} around their crossing, for all five trimers. Blue: the
energies of the two spinor levels that derive from them. Orbitals of different symmetry cross;
spinors, which all have the same symmetry in $C_{2v}$, do not. The number is the separation of the
two spinor levels where the orbitals cross.}
\label{fig:spinors}
\end{figure}

If spin--orbit coupling mixes the two orbitals, the levels that derive from them should avoid each
other. Figure~\ref{fig:spinors} shows that they do. We add the spin--orbit operator to the Fock
operator of Figure~\ref{fig:crossing} and diagonalize; each level is now a pair of spinors. In ozone
the gap between the two frontier levels is only \OrbGapO{} eV and cannot be seen on the scale of the
figure. In \ce{S3} it is \OrbGapS{} eV, in \ce{Se3} \OrbGapSe{} eV, and in \ce{Te3} \OrbGapTe{} eV.
In \ce{Po3}, with \OrbGapPo{} eV, a crossing is no longer recognizable: the lower level changes
smoothly from $2b_1$ to $4b_2$ character, and the two electrons can remain in it. In this
one-electron picture the ring opening of \ce{Po3} is no longer forbidden.

For two levels alone the gap would be the orbital matrix element $\lambda_z$ of eq~\ref{eq:me}, and
in the lighter trimers it is; in \ce{Te3} and \ce{Po3} the other valence orbitals contribute and
reduce it somewhat (Table~S22). In units of the spin--orbit constant $\zeta_{np}$ of
the atom, $\lambda_z$ is nearly the same in all five molecules, \OrbLamZetaMin{} to \OrbLamZetaMax:
the shape of the two orbitals changes little down the group, and the size of the matrix element is
set by the atom. Its value can be estimated from the local picture of Figure~\ref{fig:orbitals}.
Near the crossing $2b_1$ is the antibonding $\pi$ orbital of a
bent chain of three atoms, $(p_2-\sqrt2\,p_1+p_3)/2$ with the apex as atom 1, and $4b_2$ the
$\sigma^*$ combination $(p'_2+p'_3)/\sqrt2$ of the in-plane functions on the terminal atoms; each
terminal atom then contributes $\zeta/2\sqrt2$, and $\lambda_z=\zeta/\sqrt2=0.71\,\zeta$.

The spinor diagram is a one-electron picture. Its counterpart
for the states is the avoided crossing of the ground state with the one component of the triplet
that has its symmetry; the gap there is $2V$, with the coupling $V$ of eq~\ref{eq:V}. Where the
coupling is weak, a molecule that arrives at the crossing changes state only with a certain
probability; where it is strong, the two curves merge into a single surface with a saddle below
the crossing (Sections~\ref{sec:profiles} and \ref{sec:kinetics}).

\section{Computational Methods}
\label{sec:methods}

All calculations are all-electron calculations with the spin-free exact two-component Hamiltonian in
its one-electron form (sf-X2C-1e)\cite{Kutzelnigg2005,Ilias2007,Liu2010} and ANO-RCC basis
sets.\cite{Roos2004} The results reported are for the triple-zeta contractions $4s3p2d1f$ (O),
$5s4p2d1f$ (S), $6s5p3d1f$ (Se), $7s6p4d2f1g$ (Te), and $8s7p5d3f1g$ (Po); surfaces with the next
smaller contractions, and checks with a quadruple-zeta contraction and with a full-valence active
space, are described in the Supporting Information. Two-electron integrals are density
fitted. The programs are based on PySCF;\cite{Sun2020} the spin--orbit configuration interaction,
the treatment of the frozen core in NEVPT2, and the mean-field spin--orbit operator are our own code
and were validated separately (Supporting Information).

\emph{States without spin--orbit coupling.} The active space contains the nine valence $p$ orbitals
with twelve electrons, CAS(12,9). The orbitals are optimized in a state-averaged
CASSCF\cite{Roos1980} over $1\,^1A_1$, $2\,^1A_1$, and $^3A_2$ with equal weights. The states
$1\,^1A_1$, $2\,^1A_1$, $^3A_2$, $^3B_1$, and $^3B_2$ are then computed in the active space, and
their dynamic correlation energy is obtained from strongly contracted
NEVPT2\cite{Angeli2001,Angeli2002} with the valence $s$ orbitals correlated and all inner shells
frozen. The open-shell singlet $^1A_2$ was treated in the same way on the points around the bending
path.

\emph{Spin--orbit operator.} The one-electron part is the difference between the two-component and
the spin-free X2C core Hamiltonians. To it we add a one-center mean-field two-electron
term\cite{Hess1996} built from Breit--Pauli integrals and the spherically averaged density of the
atom. The resulting spin--orbit constants $\zeta_{np}$, obtained from the $^3P_1$ and $^1D_2$ levels
of the atoms, are 0.94, 0.93, 0.92, 0.90, and 0.84 of the experimental values\cite{NIST2024} for O
to Po. This
deficit is not caused by the operator: put into the Hartree--Fock equations of the
atoms, it reproduces the splitting of the valence $p$ spinor energies of four-component
Dirac--Coulomb--Breit calculations to within \OneAtomDevMax\% for the five chalcogen atoms and to
within \OneNobleDevMax\% for the closed $p$ shells of Ne to \OneNobleLast{}
(Tables~S17 and S18). We
attribute the deficit to the missing response of the orbitals to the coupling, since the operator
is applied in a second step; its effect on the barriers is estimated in
Section~\ref{sec:reliable}.

\emph{Spin--orbit configuration interaction.} The operator $\hat H_{\mathrm{sf}}+\Hso$ is
diagonalized in all 18\,564 determinants of twelve electrons in the eighteen active spin orbitals
with the Davidson method;\cite{Davidson1975} no selection of states is involved. Dynamic correlation
enters through a dressing of the five states $k$ listed above with their NEVPT2 correlation energies
$\delta_k$,
\begin{equation}
\hat H=\hat H_{\mathrm{sf}}+\Hso+\sum_k|k\rangle(\delta_k-\bar\delta)\langle k|+\bar\delta ,
\end{equation}
where $\bar\delta$ is the mean of the $\delta_k$. The surfaces without and with spin--orbit coupling
are thus obtained at the same level, and their difference is the effect of $\Hso$ alone. For the
control calculation of Section~\ref{sec:components} the configuration interaction was repeated with
the component of $\hat{\mathbf h}$ along the twofold axis alone and with the two perpendicular
components alone.

\emph{Surfaces.} For each molecule the energies were computed on a grid of 35 angles ($50^\circ$ to
$135^\circ$) and 25 bond lengths (steps of 0.03 \AA) of the $C_{2v}$ geometries and interpolated
with bicubic splines.
\ifnum\NDiscarded=0
The CASSCF calculations converged at all 875 points.
\else
The CASSCF calculations converged at all 875 points except \NDiscPo{} of \ce{Po3} in the corner of
the smallest angle and the shortest bonds, more than 60 \kcal{} above the lowest point of the
surface.
\fi
Minima, saddles, and the lowest points of the crossings were determined on the splines. The saddle
of the closed-shell singlet is a narrowly avoided crossing of two configurations, a kink that this
grid rounds off; the apex angle was therefore scanned through it in steps of $0.125^\circ$, and the
height of the saddle was corrected accordingly, by up to \CuspO{} \kcal{}
(Table~S12). The third
vibration, the antisymmetric stretch, was examined by displacing these points to bond lengths
$r\pm\delta$ with $\delta$ up to 0.06 \AA{} and repeating the calculation in $C_s$ symmetry.

\emph{Orbital and spinor energies.} The correlation diagrams of Figures~\ref{fig:crossing} and
\ref{fig:spinors} use the restricted open-shell Hartree--Fock density of $^3A_2$, in which $4b_2$ and
$2b_1$ hold one electron each. The eigenvalues of its spin-averaged Fock operator,
$\hat F=\hat h+\hat J[\rho]-\tfrac12\hat K[\rho]$, are the orbital energies; adding the spin--orbit
operator and diagonalizing in the two-component space gives the spinor energies.

\emph{Coupled-cluster reference.} Where every state involved has one dominant configuration (at the
ring, at the open form, and around the ring-side crossing with the triplet) the energies were
recomputed with CCSD(T), with the same Hamiltonian, basis set, and frozen core, at grid points and
with identical local fits for both methods.

\emph{Rate estimates.} Rate constants are obtained from transition-state theory over a saddle and
from nonadiabatic transition-state theory\cite{Lorquet1988,Harvey2007} through the lowest point of
a crossing, with the thermally averaged Landau--Zener\cite{Landau1932,Zener1932} probability that
the molecule changes state on its way out or on its way back and with harmonic partition
functions. The coupling is the spin--orbit matrix element at the crossing with the triplet and
the rotational coupling, averaged over a thermal distribution of $J_z$, at the crossing with the
open-shell singlet. Tunneling is estimated separately: for a crossing
with the weak-coupling formula,\cite{Delos1972,Harvey2007} which is exact for two linear
potentials, in its uniform semiclassical form with the action integrals of the computed potentials
along the normal of the seam, and for the saddle of the singlet with the one-dimensional
transmission through the computed profile along the relaxed bending path
(Section~S4).

\section{Results and Discussion}
\label{sec:results}

\subsection{Energy Profiles and Bottlenecks}
\label{sec:profiles}

\begin{figure}
\centering
\includegraphics[width=\textwidth]{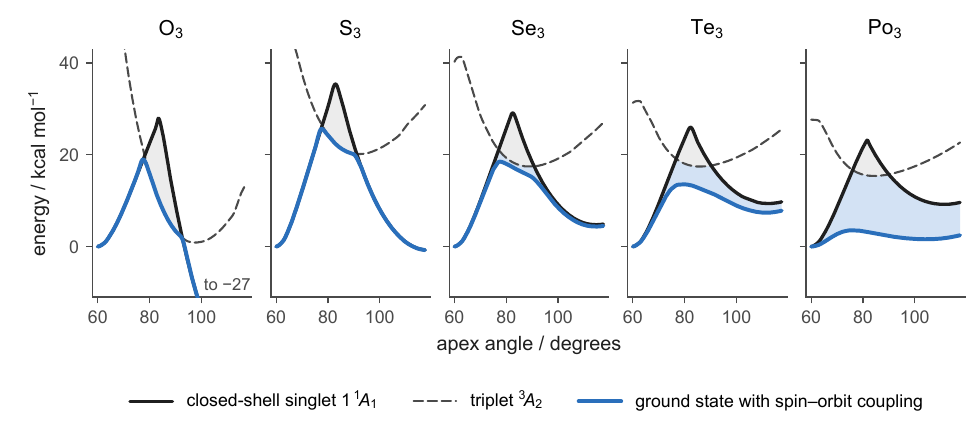}
\caption{Energy profiles of the ring opening (NEVPT2; bond length relaxed for every curve at each
angle; the lines join points computed every $2.5^\circ$ and the saddle of the singlet). Black: the
closed-shell singlet with the forbidden barrier. Dashed: the triplet; both
relative to the ring minimum of the singlet. Blue: the ground state with spin--orbit coupling,
relative to its own ring minimum. Grey area:
the part of the singlet barrier that lies above the triplet. Blue area: the further lowering by
spin--orbit coupling, relative to the lowest route without it.}
\label{fig:profiles}
\end{figure}

Figure~\ref{fig:profiles} shows the ring opening of the five trimers without and with spin--orbit
coupling. Without it the closed-shell singlet has a barrier of \BarSingPo{} to \BarSingS{} \kcal{}
above the ring, with the sharp peak expected for a forbidden reaction, and in all five molecules
the triplet crosses below the peak.

The lowering of the bottleneck relative to the forbidden saddle consists of two parts
(Table~\ref{tab:energy}). The first is the difference between the saddle and the lowest point of
the crossing with the triplet, \BypassPo{} to \BypassS{} \kcal. It is not an effect of spin--orbit
coupling, since the crossing is a feature of the surfaces without it; spin--orbit coupling only
makes a change of state possible there. The second part, from the crossing to the bottleneck of
the surface with spin--orbit coupling, is due to the coupling alone. In \ce{O3} and \ce{S3} it is
small. The coupling at the crossing, \VcmCtlO{} and \VcmCtlS{} \cm, opens a gap of \ScanGapO{} and
\ScanGapS{} \cm{} between the ground state and the triplet component that mixes with it, twice the
coupling, as expected (Figure~S2); the lower state thus passes \ScanHalfO{} and
\ScanHalfS{} \kcal{} below the crossing. For these two molecules we take the crossing itself as the
bottleneck, as in nonadiabatic transition-state theory. From \ce{Se3} on the coupling is large
enough to merge the two states into one smooth surface whose saddle lies below the crossing, by
\ExtraSe, \ExtraTe, and \ExtraPo{} \kcal{} in \ce{Se3}, \ce{Te3}, and \ce{Po3}.

\begin{table}
\caption{Ring Opening of the Chalcogen Trimers without and with Spin--Orbit Coupling (SO)$^a$}
\label{tab:energy}
\begin{tabular}{lccccccc}
\hline
 & \multicolumn{2}{c}{$E_{\mathrm{ring}}-E_{\mathrm{open}}$} & \multicolumn{3}{c}{bottleneck above the ring} & & \\
\cline{2-3}\cline{4-6}
 & without SO & with SO & saddle & crossing & with SO & $V$/\cm & $V/\zeta_{np}$ \\
\hline
\ce{O3} & \ensuremath{+}27.3 & \ensuremath{+}27.3 & 28.0 & 20.0 & 20.0 & 87 & 0.58 \\
\ce{S3} & \ensuremath{+}0.7 & \ensuremath{+}0.8 & 35.4 & 26.4 & 26.4 & 199 & 0.55 \\
\ce{Se3} & \ensuremath{-}4.8 & \ensuremath{-}4.4 & 29.1 & 21.1 & 18.5 & 948 & 0.57 \\
\ce{Te3} & \ensuremath{-}9.4 & \ensuremath{-}7.4 & 26.0 & 19.1 & 13.6 & 2037 & 0.57 \\
\ce{Po3} & \ensuremath{-}9.2 & \ensuremath{-}1.6 & 23.2 & 16.7 & 3.6 & 6013 & 0.58 \\
 
\hline
\end{tabular}

\vspace{4pt}
\parbox{0.86\linewidth}{\small $^a$Energies in \kcal{} (NEVPT2, triple-zeta basis). Saddle: of
$1\,^1A_1$; crossing: lowest point of its ring-side crossing with $^3A_2$; both belong to the
surfaces without SO. The bottleneck with SO is this crossing in \ce{O3} and \ce{S3} and the saddle
of the coupled surface in \ce{Se3} to \ce{Po3}. $V$ is the coupling between $1\,^1A_1$ and $^3A_2$
at their crossing. The entries with SO do not contain the response of the orbitals to the coupling,
which is estimated to lower the bottleneck by \CasTopAbsSe{} (\ce{Se3}), \CasTopAbsTe{} (\ce{Te3}),
and \CasTopAbsPo{} \kcal{} (\ce{Po3}); this estimate, made without dynamic correlation, is not
applied (Section~\ref{sec:reliable}).}
\end{table}

The coupling between the ground state and the triplet, evaluated directly from the wave functions,
grows by a factor of seventy from \ce{O3} to \ce{Po3}, but in units of the atomic spin--orbit
constant it is \VzetaMin{} to \VzetaMax{} throughout (Table~\ref{tab:energy}), close to the value
$\lambda_z/\sqrt2$ expected from eq~\ref{eq:V} where one configuration dominates.

\begin{figure}
\centering
\includegraphics[width=\textwidth]{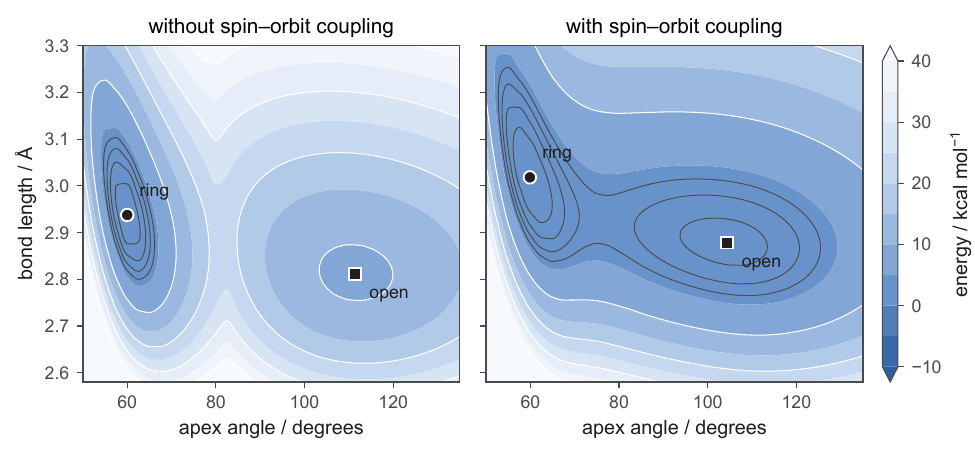}
\caption{The potential energy surface of \ce{Po3} in the coordinates apex angle and bond length
(NEVPT2), relative to the ring minimum of each surface; contour interval 5 \kcal, with thin lines
at 1, 2, 3, and 4 \kcal. Left: the closed-shell singlet without spin--orbit coupling. Right: the
ground state with spin--orbit coupling.}
\label{fig:landscape}
\end{figure}

For \ce{Po3} the surface changes qualitatively (Figure~\ref{fig:landscape}). Without
spin--orbit coupling there are two valleys, ring and open form, separated by a ridge. With
spin--orbit coupling the ridge has disappeared in our calculations, and a single broad basin
remains in which the ring is a shallow minimum, \BarSOPo{}
\kcal{} deep; the energy of the ring relative to the open form changes from \RingOpenPo{} to
\RingOpenSOPo{} \kcal. The surfaces of all five molecules are shown in Figure~S5.

\subsection{Components of the Spin--Orbit Operator}
\label{sec:components}

\begin{figure}
\centering
\includegraphics{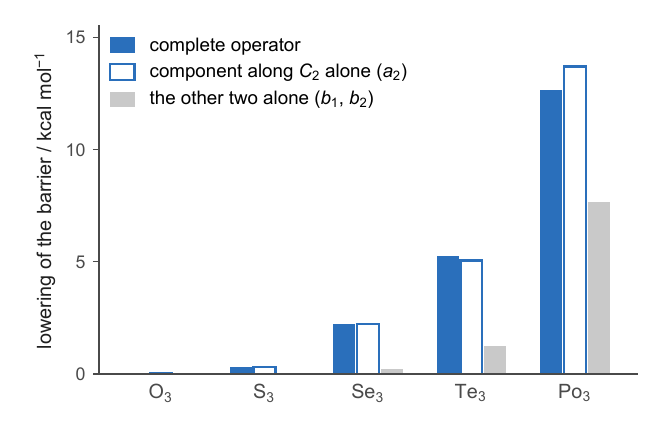}
\caption{Lowering of the barrier by the components of the spin--orbit operator, relative to the lowest
route without spin--orbit coupling (through the crossing with $^3A_2$), for the complete operator,
for its component along the twofold axis alone, and for the two perpendicular components alone;
bending profiles with the bond length relaxed for each calculation.}
\label{fig:components}
\end{figure}

Spin--orbit coupling lowers the energy of every state of a molecule with heavy atoms, so that a
lower barrier alone does not show that the mechanism proposed here is responsible. The selection
rule makes a more specific prediction: of the three components of the spin--orbit operator, only
the one along the twofold axis couples the ground state to the triplet of the crossing, and the
second part of the lowering should be due to this component. We tested
this by repeating the calculation with one part of the operator at a time
(Figure~\ref{fig:components}).

In \ce{Se3} and \ce{Te3} the prediction is fulfilled to within \CtlParDevMax{} \kcal{} (the
barriers differ by \CtlParDevSe{} and \CtlParDevTe{} \kcal). On the bending profiles
of this calculation the bottleneck lies at \CtlLSSe{} and \CtlLSTe{} \kcal{} without spin--orbit
coupling and at \CtlFullSe{} and \CtlFullTe{} \kcal{} with the complete operator; the component
along the axis alone gives \CtlParSe{} and \CtlParTe{} \kcal, the two perpendicular components alone
\CtlPerpSe{} and \CtlPerpTe{} \kcal. The small effect of the perpendicular components on the
barrier is not due to their size: from \ce{Se3} on they provide more than half of the total
spin--orbit stabilization of the ring, but they shift ring, bottleneck, and open form by similar
amounts.

In \ce{Po3} the prediction holds only qualitatively. The complete operator lowers
the bottleneck from \CtlLSPo{} to \CtlFullPo{} \kcal; the component along the axis alone gives
\CtlParPo{} \kcal, a somewhat larger lowering than the complete operator, and the two perpendicular
components alone give \CtlPerpPo{} \kcal, which is also a substantial lowering. The spin--orbit
constant of polonium is comparable to the separation of the states, so that the triplets $^3B_1$
and $^3B_2$, which the perpendicular components reach, are mixed into the ground state in second
order, and the contributions of the components are no longer additive. The component named by the
rule alone still gives a lowering of about the size found with the complete operator, but a
description by the direct coupling of two states is no longer sufficient.

\subsection{Spin--Orbit Coupling and Rotational Coupling}
\label{sec:rotation}

\begin{figure}
\centering
\includegraphics{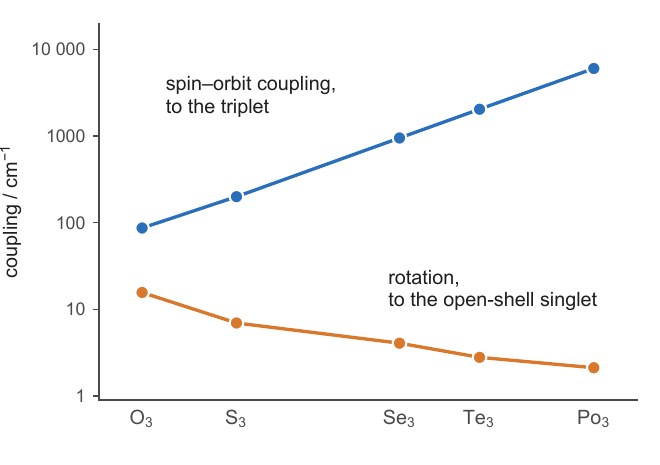}
\caption{Two couplings of the same symmetry. Blue: spin--orbit coupling between the ground state and
$^3A_2$ at their crossing. Orange: rotational coupling between the ground state and $^1A_2$ at their
crossing, root mean square at 300 K. The molecules are placed at the spin--orbit constant of their
atom on a logarithmic axis.}
\label{fig:couplings}
\end{figure}

Even without spin--orbit coupling the forbidden saddle is not the only route. The
open-shell singlet crosses the ground state \AtwoSeamO, \AtwoSeamS, \AtwoSeamSe, \AtwoSeamTe, and
\AtwoSeamPo{} \kcal{} above the ring from \ce{O3} to \ce{Po3}, about 4 to 6 \kcal{} below the
saddle, and in a rotating molecule the two states are coupled. The matrix element of $\hat L_z$
between them is close to $\hbar$ in all five molecules, in line with Figure~\ref{fig:orbitals}. The
coupling is nevertheless small, because the rotational energy is small: at 300 K the coupling of
eq~\ref{eq:V} is \VrotO{}
\cm{} in \ce{O3} and falls to \VrotPo{} \cm{} in \ce{Po3}, since the heavier molecule rotates more
slowly (Figure~\ref{fig:couplings}).

The two couplings differ by a factor of \VsoVrotO{} in ozone and change in opposite directions down
the group; in \ce{Po3} they differ by a factor of \VsoVrotPo. The rotational route is nevertheless
relevant for the light trimers. A molecule changes state at this crossing only once or twice in a
thousand passages, but the crossing lies low enough for the route to be
faster than the passage over the saddle, by factors of \RotGainO{} (\ce{O3}), \RotGainS{} (\ce{S3}),
and \RotGainSe{} (\ce{Se3}) in our estimate; in \ce{Te3} and \ce{Po3} the two are comparable.
Without spin--orbit coupling a light trimer would therefore also open its ring through the
diradical configuration. We use the two spin-conserving routes together as the reference for the
effect of spin--orbit coupling.

\subsection{Kinetics of the Ring Opening}
\label{sec:kinetics}

The thermally averaged probability of a change of spin in one
passage through the crossing is \PsingleO{} in \ce{O3} and \PsingleS{} in \ce{S3}; counting the way
out and the way back, \PhopO{} and \PhopS{} of the molecules that reach the crossing end up in the
triplet. From \ce{Se3} on the probability is practically one, and the molecule moves on a single
surface. In this description the ring opening of \ce{O3} and \ce{S3} is a spin-forbidden reaction
whose bottleneck is a crossing, and that of \ce{Se3}, \ce{Te3}, and \ce{Po3} a reaction over a
saddle.

For \ce{O3} and \ce{S3} the entry into the triplet is not yet the opening of the ring. A molecule
in the well of the diradical has to change back at a second crossing on the far side, which lies
lower, \OpenSeamO{} and \OpenSeamS{} \kcal{} above the ring, with a coupling of the same size. A
molecule that comes to rest in the well goes on to the open form; in a one-dimensional classical
model in which it keeps its energy in the reaction coordinate, only \FballO{} (\ce{O3}) and
\FballS{} (\ce{S3}) of the molecules go on (Table~S5). The rate constants that we
report for these two molecules are those of the first entry into the triplet well.

Rate constants from transition-state theory on these surfaces are collected in the Supporting
Information (Table~S8, Figure~S1). We consider their ratios more
reliable than their absolute values: the partition functions are harmonic, the barriers carry the
errors discussed in Section~\ref{sec:reliable}, and one \kcal{} is a factor of five at room
temperature. Relative to the two spin-conserving routes together, the route that spin--orbit
coupling opens is faster by factors of \SpeedupO{} (\ce{O3}), \SpeedupS{} (\ce{S3}), \SpeedupSe{}
(\ce{Se3}), \SpeedupTe{} (\ce{Te3}), and \SpeedupPo{} (\ce{Po3}) at 300 K; on these surfaces the
ring of \ce{Po3} would open after \HalfSfPo{} without spin--orbit coupling and after \HalfBestPo{}
with it. The spin-conserving reference assumes an
adiabatic passage over the saddle of the singlet, where the two $^1A_1$ states are only
\SadGapMin{} to \SadGapMax{} eV apart; it is an upper bound in this respect, and the factors above
are lower bounds.

Ozone requires separate consideration, because for oxygen atoms and a barrier this narrow tunneling
cannot be neglected. Chen and Hu computed the rate of the ring opening with multidimensional
tunneling and obtained a half-life of 0.05 s at 300 K.\cite{Chen2011} Their barrier of 21.1 \kcal{}
is not the saddle of the ground state, which they find at 25.8 \kcal, but its crossing with the
open-shell singlet, on the assumption that the reaction follows the lower of the two surfaces; on
our surfaces the saddle lies at \BarSingO{} and this crossing at \AtwoSeamO{} \kcal. The
selection rule bears on this assumption: the two singlets differ by the symmetry $a_2$, and the
only coupling between them is the rotational one, with a thermally averaged probability of
\ProtO{} for a change of state at 300 K. The triplet offers
a more favorable route in both respects: it crosses the ground state lower, \BarCrossO{} \kcal{}
above the ring, and it is reached with a probability of \PhopO{} that does not depend on rotation.
Neither advantage rests on the accuracy of the computed barriers: by Hund's rule the triplet lies
below the open-shell singlet of the same configuration and therefore crosses the ground state at
lower energy, and the two probabilities differ by a factor of about \PhopOverProtO. One-dimensional estimates of tunneling through the
computed potentials (Table~S9) raise the rate of entry into the triplet by a factor
of about \KappaTunO, which brings the modeled median time to this entry from \HalfBestO{} to
\HalfTunO, the rotational route by \KappaRotO, and the passage over the saddle by \KappaWkbO; the entry into the triplet then stays ahead of the other two together by a
factor of about \TunLeadO. Nearly all of this flux, \ShareBelowO\%, enters below the crossing, and
in the one-dimensional model \FtunO{} of the entering molecules go on to the open form. Our modeled time lies between the half-lives of
0.05 and 0.55 s that multidimensional tunneling gave with two other sets of electronic
energies;\cite{Chen2011,Jiang2025} all three values are predictions whose accuracy is limited by
that of the barrier. Our estimates suggest that the reaction proceeds through the triplet and that
the rate-limiting step is a change of spin.

\subsection{Accuracy of the Calculations}
\label{sec:reliable}

The basis set matters for the absolute barriers, which rise by 2 to 7 \kcal{} from the double- to
the triple-zeta contraction, but less for the differences on which the argument rests: the distance
from the singlet saddle to the bottleneck with spin--orbit coupling changes by at most 1.5 \kcal{}
(Table~S3). In a quadruple-zeta contraction, tested on the bending profiles of the
control calculation, the barriers rise further by at most \ChkQzBarMax{} \kcal, and the
differences between them change by at most \ChkQzDiffMax{} \kcal; the component of the operator
along the twofold axis alone again gives the barrier of the complete operator to within
\ChkQzParMax{} \kcal{} in \ce{Se3} and \ce{Te3} (Table~S4).

The comparison with CCSD(T) shows where the error of NEVPT2 is largest (Table~S20).
Relative to the ring, CCSD(T) places the open form, the triplet, and the singlet near the crossing
lower by \RefShiftLoO{} to \RefShiftHiO{} \kcal{} in \ce{O3}, by \RefShiftLoS{} to \RefShiftHiS{}
\kcal{} in \ce{S3}, and by \RefShiftLoSe{} to \RefShiftHiSe{} \kcal{} in \ce{Se3}; for \ce{Te3} and
\ce{Po3} the two methods agree to about 1 \kcal. NEVPT2 with this active space thus overstabilizes
the ring of the lighter trimers. The literature points the same way: for \ce{S3} the ring lies
\RefRingOpenNevS{} \kcal{} above the open form with NEVPT2, \RefRingOpenCCS{} \kcal{} with CCSD(T),
and $+4.4\pm0.5$ \kcal{} in the composite coupled-cluster treatment of Peterson et
al.\cite{Peterson2006} Most of this error belongs to the active space: with the three valence $s$
orbitals included, CAS(18,12), the ring of \ce{S3} lies \ChkFvRingOpenS{} instead of
\CtlRingOpenS{} \kcal{} above the open form on the bending profiles of the control calculation and
its singlet barrier is lower by \ChkFvBarShiftAbsS{} \kcal, while the differences between the
routes change by at most \ChkFvDiffMax{} \kcal{} in all five molecules
(Table~S4). Density functionals
give, on average, barriers of 26, 17, and 10 \kcal{} above the open form for \ce{S3}, \ce{Se3}, and \ce{Te3},\cite{Goddard1999} where
our saddles lie at \BarOpenS, \BarOpenSe, and \BarOpenTe{} \kcal; the trend down the group is the
same, but the absolute height depends strongly on the method. Barriers
measured from the ring, and with them the absolute rate constants, carry these errors. In the
differences between routes the common reference, the ring, drops out, but errors that depend on the
state and on the geometry can remain; the saddle of the closed-shell singlet itself, where two
configurations have equal weight, cannot be checked with coupled-cluster theory.

The shallow minimum that remains at the ring of \ce{Po3} is the result that depends most on the
details of the calculation (Tables~S3 and S16). On the bending
profiles of the control calculation its depth is \SensRefPo{} \kcal; it is \SensThreePo{} and
\SensNinePo{} \kcal{} when three or nine states instead of five are dressed with their NEVPT2
energies, falls to
\SensExpPo{} \kcal{} when the operator is scaled to the experimental spin--orbit constant of the
atom, is \ChkQzFullPo{} \kcal{} in the quadruple-zeta basis and \ChkFvFullPo{} \kcal{} with the
full-valence active space, and is only about 1 \kcal{} without NEVPT2. The ring of \ce{Po3} with
spin--orbit coupling is a minimum in all these calculations, but its depth is known only to within
a few \kcal.

The spin--orbit operator itself agrees with four-component calculations to within one percent for
the atoms of all five rows (Section~\ref{sec:methods}), so that neither the Breit interaction nor
the mean-field form of the two-electron term limits the accuracy. The weaker point is that the
operator acts in a second step, on orbitals that cannot respond to it. To estimate this response,
we repeated the complete configuration interaction in the valence $p$ space with spinors from a
two-component Hartree--Fock calculation that contains the operator, at the ring, the saddle, and
the open form (Section~S11). With the $p$ primitives
released, the response of the orbitals lowers the saddle relative to the ring by \CasTopAbsSe,
\CasTopAbsTe, and \CasTopAbsPo{} \kcal{} in \ce{Se3}, \ce{Te3}, and \ce{Po3}
(Table~S19), about what the scaling of the operator to the experimental constants of
the atoms had suggested. The two-step treatment thus probably overestimates the barriers of the
heavy trimers with spin--orbit coupling by about one \kcal; the minimum at the ring of \ce{Po3}
would then be about \CasBarEstPo{} instead of \BarSOPo{} \kcal{} deep, and the route opened by
spin--orbit coupling would be faster in \ce{Te3} and \ce{Po3} by a factor of about five at room
temperature. Because the estimate concerns one effect at three geometries and contains no dynamic
correlation, we do not apply it to Table~\ref{tab:energy}, and it is not an error bar.

Finally, the surfaces are restricted to $C_{2v}$. The third vibration, the antisymmetric stretch,
cannot remove the forbiddenness, which belongs to every planar geometry, but it could lower the
bottlenecks. In our calculations it does not: displacements along it raise the energy at the saddle
of the singlet, on
the seam with the triplet, and at the saddle with spin--orbit coupling in all five molecules
(Table~S15), so that the bottlenecks of the two-dimensional surfaces are bottlenecks
in all three dimensions.

\section{Scope of the Selection Rule: The Pnictogen Tetramers}
\label{sec:beyond}

Molecules of more than three atoms can leave the symmetric path by a vibration, and the route
through the triplet then competes with the distorted singlet path. The selection rule does not
decide this competition; it indicates for which forbidden reactions of heavy elements a route
through the triplet exists, and the local argument of Figure~\ref{fig:orbitals} indicates how large
the coupling is likely to be.

The tetrahedral pnictogens serve as an example. The dissociation \ce{E4 -> 2 E2} along the
least-motion path of $D_{2d}$ symmetry is forbidden by the crossing of a $b_1$ with a $b_2$ orbital,
and $b_1\otimes b_2=a_2$ is again the rotation about the principal axis. The rule is fulfilled, but
the matrix element is only \EfourRatioMin{} to \EfourRatioMax{} of the atomic spin--orbit constant,
against \OrbLamZetaMin{} to \OrbLamZetaMax{} in the trimers (Tables~S23 and
S24): the two orbitals of the tetramer are not related by a simple quarter turn on
each atom, and the contributions of the four atoms partly cancel. The symmetry of a rotation is
thus a necessary condition; a large effect requires in addition that a heavy atom contributes to
the two orbitals $p$ functions that the rotation converts into each other.

\begin{figure}
\centering
\includegraphics[width=\textwidth]{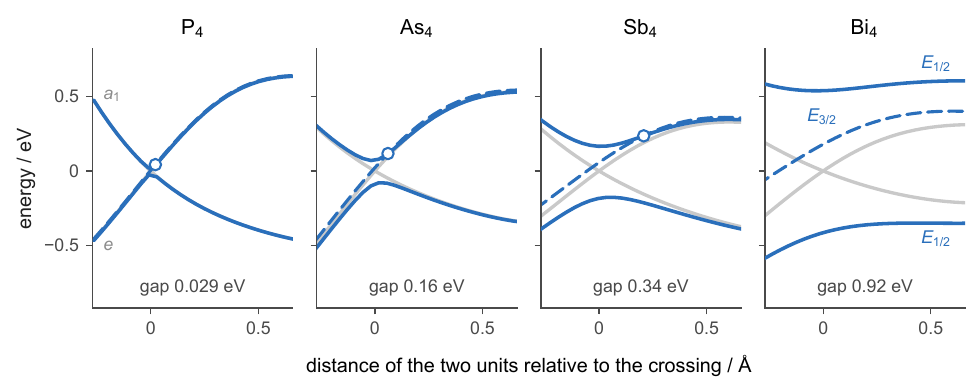}
\caption{Levels that mix and a level that does not, in the four tetramers along the $D_{2d}$ path
of \ce{E4 -> 2 E2}. Grey: an $a_1$ and an $e$ orbital of the occupied valence shell, which cross.
Blue: the three spinor levels that derive from them. The two levels of species $E_{1/2}$ (solid)
avoid each other, with the gap given; the level of species $E_{3/2}$ (dashed), the second component
of $e$, has no partner of its symmetry and stays unmixed. Circles: points at which it crosses an
$E_{1/2}$ level. Energies relative to the crossing of the two orbitals.}
\label{fig:protected}
\end{figure}

The tetramers also allow a test of the converse statement of the rule, which the trimers cannot
provide. In $C_{2v}$ the product of any two different representations is a rotation, and the double
group has a single double-valued irreducible representation, so that no crossing is protected. In
$C_s$ and $C_2$, the groups in which the electrocyclic reactions are usually
analyzed,\cite{LH1965,WH1969} the two representations also differ by a rotation, so that there
the rule always admits the coupling and only its size is in question. In
$D_{2d}$ the rotations span only $a_2$ and $e$, and there are two double-valued irreducible
representations, $E_{1/2}$ and $E_{3/2}$. The spinors of an $a_1$ or $a_2$ orbital belong to the
first, those of a $b_1$ or $b_2$ orbital to the second, and a degenerate $e$ orbital gives one
level of each kind. Among the occupied valence orbitals an $a_1$ and an $e$ level cross on the same
path (Figure~\ref{fig:protected}). With spin--orbit coupling the two levels of species $E_{1/2}$
avoid each other, with a gap that grows from \ProtGapP{} eV in \ce{P4} to \ProtGapBi{} eV in
\ce{Bi4} and is close to the spin--orbit constant of the atom (\ProtGapZetaMin{} to
\ProtGapZetaMax{} $\zeta_{np}$). The level of species $E_{3/2}$, the other component of $e$, has no
partner of its symmetry; it mixes with neither of the two and follows the course of the $e$
orbital. None of these crossings makes a reaction forbidden, since the orbitals involved are
occupied. We show them because they display both cases in one molecule and with the same
spin--orbit constant: which of the two occurs is determined by symmetry and not by the size of the
coupling.

The same two criteria can be applied to other forbidden reactions. If the product of the crossing
orbitals is not a rotation, as for a \emph{gerade} and an \emph{ungerade} orbital, spin--orbit
coupling provides no route through the triplet, however heavy the atoms; the stretching of the
central bond of a [2.2.2]propellane, where a $\sigma$ and a $\sigma^*$ orbital change places that
differ by the symmetry of a translation along the bond,\cite{Stohrer1972} would be a reaction of
this kind. If the product is a rotation, the size of the effect depends on whether the two orbitals
contain $p$ functions on the same heavy atom that differ by a quarter turn. In the idealized
suprafacial combination of two $\pi$ bonds they do not, and we expect the effect to be small; for
the lone pair and the empty $p$ orbital in the least-motion dimerization of heavy carbene
analogues they do, and we expect spin--orbit coupling to lower the forbidden barrier
substantially. These expectations have not been tested by calculations here.

\section{Conclusions}

We have examined how spin--orbit coupling affects a reaction that is forbidden by orbital symmetry.
The two crossing orbitals define a diradical configuration whose triplet lies below the forbidden
saddle, in the two-orbital model and in all five trimers studied. Spin--orbit coupling can connect
this triplet to the ground state if, and only if, the two orbitals differ by the symmetry of a
rotation, and it then acts through the operator component of that symmetry. In the ring opening of
the chalcogen trimers this is the component along the twofold axis, which alone gives the barrier
of the complete operator to \CtlParDevMax{} \kcal{} in \ce{Se3} and \ce{Te3}. In \ce{Po3} the spin--orbit
constant is comparable to the separation of the states; the contributions of the components are
then no longer additive, and a first-order selection rule does not describe the situation
completely. In \ce{O3} and \ce{S3} the named component allows a change to the triplet at the
crossing with low probability, so that the reaction becomes a spin-forbidden one; from \ce{Se3} on
it turns the crossing into the saddle of a single surface, and in \ce{Po3} only a shallow minimum,
whose depth depends on the method, remains at the ring. The rotation of the molecule obeys the
same selection rule but gives a much weaker coupling, and our estimates suggest that the short
lifetime predicted for cyclic ozone is determined by its triplet rather than by the open-shell
singlet. Two spinor levels that have no representation in common do not mix, as the tetramers
show. For molecules with heavy atoms a selection rule for spin--orbit coupling thus appears to be
a useful supplement to the orbital-symmetry rules. The surfaces used here are restricted to
$C_{2v}$, and the kinetic estimates are simple; calculations of the nuclear dynamics and tests on
larger molecules will be needed to show how general these conclusions are.

\begin{acknowledgement}
The author was supported by the Deutsche Forschungsgemeinschaft (DFG, German Research Foundation)
under Project 535298924.
\end{acknowledgement}

\section*{Use of AI Tools}

The large language models Claude Opus 5.5 and Claude Fable 5.2 (Anthropic), operated as coding
agents (Claude Code), were used extensively in all stages of this work: in developing the argument,
in writing and testing the programs, in running and evaluating the calculations, in preparing the
figures, and in drafting and revising the text of the article and of the Supporting Information.
The author directed the work and takes full responsibility for its content. The programs written
in this way were validated against independent implementations and limiting cases as described in
the Supporting Information.

\begin{suppinfo}
Basis sets; atomic calibration and four-component tests of the spin--orbit operator; barriers at
three levels of theory in two basis sets; checks with a quadruple-zeta basis and with a
full-valence active space; stationary points, crossings, and the estimates of rates
and tunneling; dense scans through the crossing of the states in \ce{O3} and \ce{S3} and through
the saddle of the singlet; profiles and surfaces of all five molecules; control calculation;
antisymmetric stretch; sensitivity of the barriers; coupled-cluster reference; open-shell singlet;
orbital and spinor energies; tetramers; validation of the programs (PDF)

Programs, raw results of all calculations, plain-text data files, and the scripts that produce
every number, table, and figure (ZIP)
\end{suppinfo}

\bibliography{refs}

\begin{tocentry}
\includegraphics{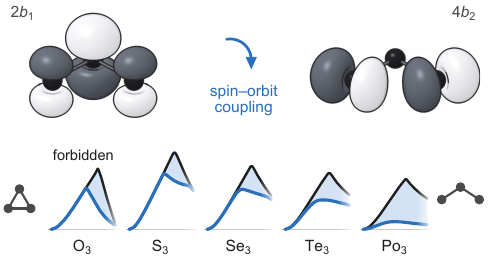}
\end{tocentry}

\end{document}